\documentclass[12pt]{article}
\global\arraycolsep=2pt
\title{\hfill {\small OKHEP--02--08}\\
Dark Energy as Evidence for Extra Dimensions}
\author{Kimball A. Milton\\
Department of Physics and Astronomy, \\University of Oklahoma,
Norman, OK 73019 USA}
\begin{document}

\maketitle

\begin{abstract}
It is argued that fluctuations of quantum fields in four-dimensional space
do not give rise to dark energy, but are rather a negligible contribution
to dark matter.  By (relativistic) dark matter we mean that the relation
between pressure and energy density is $p=\frac13 u$, while dark energy
is characterized by $p=-u$.
A possible source of dark energy are the fluctuations in
quantum fields, including quantum gravity, inhabiting extra compactified
dimensions.  These fluctuations have been computed for some simple geometries,
such as $S^2$, $S^4$, and $S^6$.
 If the extra dimensions are too small,
they would give rise to a dark energy larger than that observed, whereas
if they are too large they would be in conflict with experimental
tests of Newton's law. This notion
suggests that
the size of the extra dimensions is of order 100 $\mu$m.
If the limit on the size of extra dimensions becomes lower than this bound,
extra dimensions probably do not exist, and another source for cosmological
dark energy will have to be found.
\end{abstract}

\section{Introduction}
It has been appreciated for many years that there is an apparently fundamental
conflict between quantum field theory and the smallness of the cosmological
constant \cite{weinberg}.
  This is because the zero-point energy of the
quantum fields (including gravity) in the universe should give rise
to an observable cosmological vacuum energy density,
\begin{equation}
u_{\rm cosmo}\sim{1\over L_{\rm Pl}^4},
\end{equation}
where the Planck length is
\begin{equation}
L_{\rm Pl}=\sqrt{G_N}=1.6\times 10^{-33}\,\mbox{cm}.
\end{equation}
(We use natural units with $\hbar=c=1$.  The conversion factor is
$\hbar c \simeq 2 \times 10^{-14}\,\mbox{GeV\,cm}$.)  This means that the
cosmic vacuum energy density would be
\begin{equation}
u_{\rm cosmo}\sim 10^{118} \mbox{ GeV\,cm}^{-3},
\label{ccprob}
\end{equation}
which is 123 orders of magnitude larger than the critical mass density
required to close the universe:
\begin{equation}
\rho_c={3H_0^2\over8\pi G_N}=1.05\times 10^{-5}h_0^2 \,\mbox{GeV\,cm}^{-3},
\end{equation}
in terms of the dimensionless Hubble constant, 
$h_0=H_0/100 \; \mbox{km\,s}^{-1}\mbox{Mpc}^{-1}$. 
~From relativistic covariance 
the cosmological vacuum energy density must be the $00$ component 
of the expectation value of the energy-momentum tensor, 
which we can identify with the cosmological constant:
\begin{equation}
\langle T^{\mu\nu}\rangle =-u g^{\mu\nu}=-{\Lambda\over8\pi G}g^{\mu\nu}.
\end{equation}
[We use the metric with signature $(-1,1,1,1)$.]
Of course this is absurd with $u$ given by Eq.~(\ref{ccprob}), which would
have caused the universe to expand to zero density long ago.

For most of the past century, it was the prejudice of theoreticians that
the cosmological constant was exactly zero, although no one could give
a convincing argument.  Recently, however, with the new data gathered
on the brightness-redshift relation for very distant type Ia supernov\ae\
\cite{sn}, corroborated by the balloon observations of the
anisotropy in the cosmic microwave back\-ground \cite{cmb}, it
seems clear that the cosmological constant is near the critical value, or 
\begin{equation}
\Omega_\Lambda=\Lambda/8\pi G\rho_c \simeq 0.6-0.7.
\end{equation}
It is very hard to understand how the cosmological
constant can be nonzero but small.

\section{Quantum Fluctuations}
We here present a plausible scenario for understanding this puzzle.
It seems quite clear that vacuum fluctuations
in the gravitational and matter fields in flat Minkowski space give
a zero cosmological constant. For example, we can consider fluctuations
of conformal matter in $R\times S^3$, which give rise to the following
forms of the energy and free energy at high and low temperature \cite{entropy}
\begin{eqnarray}
F&\sim&-\frac1a a_4(2\pi aT)^4,\quad E\sim \frac1a 3a_4(2\pi aT)^4,
\quad aT\gg1,\nonumber\\
F&=&E=\frac{a_0}a, \quad aT\ll1,
\end{eqnarray}
where $a$ is the radius of $S^3$.
Here, for example, $a_4=1/48$ and $a_0=3/16$ for ${\cal N}=4$ SUSY.
[It should be noted that at present
 the universe is in the high temperature regime;
since inflation $aT\sim 10^{29}.$]
Either regime corresponds to a relation typical of radiation,
\begin{equation}
p=-\frac{\partial}{\partial V}F=\frac13u,\quad
u=\frac{E}{V}, \quad V=2\pi^2 a^3.
\end{equation}

 Other vacuum fluctuation phenomena are also
unlikely to contribute to the cosmological constant.  For example, 
fluctuations of quark and gluon fields inside a hadronic bag of radius $R$
give a zero-point energy of roughly \cite{bag}
\begin{equation} 
E_{\rm ZPE}\sim{0.7\over R}.
\end{equation}
Johnson's model of the QCD vacuum \cite{johnson} as consisting of a sea
of virtual bags might suggest then a corresponding energy density
\begin{equation}
u_{\rm QCD}\sim{E_{\rm ZPE}\over{4\pi\over 3}R^3}\sim10^{39}\mbox{GeV\,cm}^{-3}
\end{equation}
if $R=0.5\, \mbox{fm}$, some 44 orders of magnitude too large.
  Yet this is
surely an unreasonable inference:  Rather $E_{\rm ZPE}$ is absorbed into a
renormalization of QCD parameters, as a contribution, for example, to the
masses of the observed hadrons.  Thus, we expect quite confidently that there
is no QCD vacuum gravitational effect.\footnote{After mentioning the
QCD zero-point contribution to the cosmological constant, Weinberg in
1989 \cite{weinberg} pointed out that the classical vacuum energy resulting
from spontaneous symmetry breaking need not give rise to an effective
cosmological constant at the present era.}

\section{Extra Dimensions}
On the other hand, since the work of Kaluza and Klein
it has been an exciting possibility that there exist extra dimensions
beyond those of Minkowski space-time.  
Why do we not experience those dimensions?  
The simplest possibility seems to be that those  extra dimensions
are curled up in a space $\cal S$ of size $a$, 
smaller than some observable limit.

Of course, in recent years, the idea of extra dimensions has become
much more compelling.  Superstring theory  requires at least 10 dimensions,
six of which must be compactified, and the putative M theory, supergravity,
is an 11 dimensional theory.
 Perhaps, if only gravity experiences the
extra dimensions, they could be of macroscopic size.
 Various scenarios
have been suggested \cite{ed}.

Macroscopic extra dimensions imply deviations from Newton's law at such
a scale.  Two years ago, millimeter scale deviations seemed plausible, and
many theorists hoped that the higher-dimensional world was on the brink
of discovery. Experiments were initiated \cite{gravexp}.
Recently,
the results of the first definitive experiment have appeared \cite{defexp},
which indicate
no deviation from Newton's law down to 200 $\mu$m.  
This poses a
serious constraint for model-builders.\footnote{We might also mention
short distance constraints on Yukawa-type corrections to the gravitational
potential coming from Casimir measurements themselves.
See Ref.~\cite{bmm}.  Most recently, stringent limits have now appeared
for Yukawa forces with ranges between 200 and 500 microns \cite{long}.}

It seems to be commonly believed that submillimeter tests of gravity
put no limits on the size of extra dimensions if $N>2$.  This is because
of the relation of the size $R$ of the extra dimensions in the ADD
scheme to the fundamental $4+N$ gravity scale $M$ \cite{addmore}:
\begin{equation}
R\sim{1\over M}\left(M_{\rm Pl}\over M\right)^{2/N},
\end{equation}
where $M_{\rm Pl}=1/L_{\rm Pl}=1.2\times 10^{16}$ TeV is the usual
Planck mass.  Moreover, the supernova limits on ADD extra dimensions
(due to production of Kaluza-Klein gravitons)
become rapidly smaller with increase in $N$ \cite{SNconstraints}:
\begin{eqnarray}
N=2:\quad R&<&0.9\times 10^{-4}\,\mbox{mm},\\
N=3:\quad R&<&1.9\times 10^{-7}\,\mbox{mm}.
\end{eqnarray}
(Cosmological constraints are even stronger \cite{fairbairn}, but are less
certain.)
Thus direct tests of Newton's law are not competitive.  However, 
as we will see, the resulting
Casimir contribution to the cosmological constant would be enormous for
such small compactified regions, and it would seem impossible to naturally
resolve this problem.

The situation at first glance seems rather different with the RS scenario.  
In the original
scheme, gravity is localized in the ``Planck brane,'' while the standard-model
particles are confined to the ``TeV brane.''  As a consequence, it might appear
that the quantum fluctuations of both brane and bulk fields are negligible
\cite{goldroth}.  It has been stated that the cosmological constant becomes
exponentially small as the brane separation becomes large \cite{tye}.
However, this is at the ``classical level,'' without bulk fluctuations;
explicit considerations show that quantum effects give rise to a large
cosmological constant, of order of that given by Eq.~(\ref{ccprob}), 
unless an appeal is made to fine tuning \cite{odintsov}.
Moreover, if the scenario is extended so that the world
brane contains compactified dimensions in which gravity lives \cite{chako},
the constraints we deduce here directly apply.

Here we propose that a very tight constraint indeed emerges if we recognize
that compact dimensions of size $a$ necessarily possess a quantum
vacuum or Casimir energy of order $u(z)\sim a^{-4}$.  
  These can be calculated in simple
cases.  Appelquist and Chodos \cite{ac}
found that the Casimir energy
for the case of scalar field on a circle, ${\cal S}=S^1$, was
\begin{equation}
u_C=-{3\zeta(5)\over64\pi^6a^4}=-{5.056\times10^{-5}\over a^4},
\end{equation}
which needs only to be multiplied by 5 for graviton fluctuations.
The general case of scalars on ${\cal S}=S^N$, $N$ odd, was considered
by Candelas and Weinberg \cite{cw}, who found that the Casimir energy
was positive for $3\le N\le 19$, with a maximum at $N=13$ of
$u_C=1.374\times 10^{-3}/a^4$. 
 The even dimensional case was much
more subtle, because it was divergent.  Kantowski and Milton \cite{km}
showed that the coefficient of the logarithmic divergence was unique,
and adopting the Planck length as the natural cutoff, found
\begin{equation}
S^N, \,\,N \mbox{ even}: \quad u^N_C={\alpha_N\over a^4}\ln{a\over L_{\rm Pl}},
\end{equation} 
but $\alpha_N$ was always negative for scalars.
In a second paper \cite{km2} we extended the analysis to vectors, tensors,
fermions, and to massive particles, among which cases positive values of the
(divergent) Casimir energy could be found.  Some representative results for
massless particles are shown in Table 1. In an unsuccessful 
 attempt to find stable configurations, the analysis was extended
to cases where the internal space was the product of spheres \cite{bkm}.

\begin{table}
\centering
\begin{tabular}{lcccc}
$\cal S$&Gravity&Scalar&Fermion&Vector\\
\hline
$S^1$&$-2.53\times10^{-4}$&$-5.06\times10^{-5}$&$2.02\times10^{-4}$&
---\\
$S^1$&$2.37\times10^{-4}$&$4.74\times10^{-5}$&$-1.90\times10^{-4}$&
---\\
$S^2$&$1.70\times10^{-2}$&$-8.04\times10^{-5}$&$-7.94\times10^{-4}$&
$-8.04\times10^{-5}$\\
$S^3$&---&$7.57\times10^{-5}$&$1.95\times10^{-4}$&---\\
$S^4$&$-0.489$&$-4.99\times10^{-4}$&
$-6.64\times10^{-3}$&$1.21\times10^{-2}$\\
$S^5$&---&$4.28\times10^{-4}$&$-1.14\times10^{-4}$&---\\
$S^6$&$5.10$&$-1.31\times10^{-3}$&$-3.02\times10^{-2}$&$4.90\times10^{-2}$\\
$S^7$&---&$8.16\times10^{-4}$&$5.96\times10^{-5}$&---\\
\end{tabular}
\caption{The Casimir energy for $M^4\times {\cal S}$ is tabulated for various
field types in the compact geometry $\cal S$.  We write $u=[\alpha
\ln(a/L_{\rm Pl})+\beta]a^{-4}$, and give $\alpha$ for even internal dimension
and $\beta$ for odd, where $\alpha=0$. For $S^1$ the
first entry denotes untwisted 
(periodic) while the second denotes twisted (antiperiodic) boundary conditions.
 The entries marked with dashes
have not been calculated.}
\end{table}

It is important to recognize that these Casimir energies correspond to a
cosmological constant in our $3+1$ dimensional world, not in the extra
compactified dimensions or ``bulk.''  They constitute an effective source
term in the 4-dimensional Einstein equations.  
Note that because the scale $a$ makes no reference to four-dimensional
space, the total free energy of the universe (of volume $V$)
arising from this source
is $F=V u_c$, so as required for dark energy or a cosmological constant,
\begin{equation}
p=-\frac{\partial}{\partial V}F=-u_c,\quad T^{\mu\nu}=-u_cg^{\mu\nu}.
\end{equation}

The goal, of course, in all these investigations was to include graviton
fluctuations.  However, it immediately became apparent that
the results
were gauge- and reparameterization-depen\-dent unless the DeWitt-Vilkovisky
formalism was adopted \cite{vilk}.
 This was an extraordinarily difficult
task. Among the early papers in which the unique effective action
is given in simple cases we cite Ref.~\cite{UAE}.
 Only in 2000 did the general analysis for gravity 
appear, with results
for a few special geometries \cite{chokan}.
Cho and Kantowski obtain the unique
divergent part of the effective action for ${\cal S}=S^2$, $S^4$, and $S^6$,
as polynomials in $\Lambda a^2$. (Unfortunately, once again, they are unable to 
find any stable configurations.)

 The results are also shown in Table 1,
for $\Lambda a^2\sim G/a^2\ll1$.
It will be noted that
graviton fluctuations dominate matter fluctuations,
except in the case of a large number of matter fields in a small number
of dimensions.
 Of course, it would be very interesting to know the
graviton fluctuation results for odd-dimensional
spaces, but that seems to be a more difficult calculation; it is far easier to
compute the divergent part than the finite part, which is all there is in
odd-dimensional spaces.

These generic results may be applied to recent popular scenarios.  For example,
in the ADD scheme  only gravity propagates in the bulk, while the
RS approach  has other bulk fields in a single extra dimension.

Let us now perform some simple
estimates of the cosmological constant in these models.  The data suggest a
positive cosmological constant, so we can exclude those cases where the
Casimir energy is negative.  For the odd $N$ cases, where the Casimir energy
is finite, let us write
\begin{equation}
{\cal S}=S^N, \quad N \,\mbox{odd}:\quad
u_C^N={\beta_N\over a^4}, 
\end{equation}
so merely requiring that this be less than the critical density $\rho_c$
implies ($\beta>0$)
\begin{equation}
a>\beta^{1/4}h_0^{-1/2} 67\, \mu\mbox{m}\approx \beta^{1/4} 80\,
 \mu\mbox{m},
\label{betalim}
\end{equation}
taking $h_0=0.7$ (with about a 10--20\% uncertainty). 
 As seen in Table 2 these lower limits (for a single species) are still
an order of magnitude below the experimental upper limit.

\begin{table}
\centering
\begin{tabular}{lcccc}
$\cal S$&Gravity&Scalar&Fermion&Vector\\
\hline
$S^1$ (u)&*&*&9.5 $\mu$m&---\\
$S^1$ (t)&9.9 $\mu$m&6.6 $\mu$m&*&---\\
$S^2$&84 $\mu$m&*&*&*\\
$S^3$&---&7.5 $\mu$m&9.5 $\mu$m&---\\
$S^4$&*&*&*&77 $\mu$m\\
$S^5$&---&11.5 $\mu$m&*&---\\
$S^6$&350 $\mu$m&*&*&110 $\mu$m\\
$S^7$&---&13.5 $\mu$m&7.0 $\mu$m&---\\
\end{tabular}
\caption{The lower limit to the radius of the compact dimensions
deduced from the requirement that the Casimir energy not exceed
the critical density. The numbers shown are for a single species of
the field type indicated.  The dashes indicate cases where the Casimir
energy has not been calculated, while
 asterisks indicate (phenomenologically
excluded) cases where the Casimir energy is negative.}
\end{table}

Much tighter constraints appear if we use the divergent results for even
dimensions.  We have the inequality ($\alpha>0$)
\begin{equation}
a>[\alpha\ln(a/L_{\rm Pl})]^{1/4}80\,\mu\mbox{m},
\end{equation}
where we can approximate $(\ln a/L_{\rm Pl})^{1/4}\approx2.9$.
Again results are shown in Table 2,
which  rules out all but
 one of the gravity 
cases ($S^2$) given by Cho and Kantowski. For matter
fluctuations only, excluded are $N>14$ for a single
vector field and $N>6$ for a single tensor field.  (Fermions always have
a negative Casimir energy in even dimensions.)
Of course, it is possible to achieve cancellations by including
various matter fields and gravity.

 In general the Casimir energy is obtained
by summing over the species of field which propagate in the extra dimensions,
 \begin{equation}
 u_{\rm tot}={1\over a^4}
\sum_i\left[\alpha_i
\ln(a/L_{\rm Pl})+\beta_i\right]\approx{\beta_{\rm eff}\over a^4},
\end{equation}
which leads to a lower limit according to Eq.~(\ref{betalim}).
Presumably, if  exact supersymmetry 
held in the extra dimensions (including
supersymmetric boundary conditions), the Casimir energy would vanish, but
this  would seem to be difficult to achieve with {\em large\/} extra
dimensions  (1 mm corresponds to $2\times 10^{-4}$ eV.)

That there is a correlation 
between the currently favored value of the cosmological constant and
submillimeter-sized extra dimensions has been noted qualitatively 
before \cite{before}.

\section{Conclusions}

We have proposed the following scenario to explain the predominance
of dark energy in the universe.
\begin{itemize}
\setlength{\itemsep}{10pt}
\item Quantum fluctuations of gravity/matter fields in extra dimensions
give rise to a dark energy, or cosmological constant, $\propto1/{a^4}$
where $a$ is the size of the extra dimensions.
\item The dark energy will be too large unless $a>10-300$ $\mu$m.
\item Laboratory (Cavendish) tests of Newton's law require $a<200$ $\mu$m.
\item Thus, we may be on the verge of discovery of extra dimensions,
or
\item Extra dimensions do not exist and dark energy has another origin.
\end{itemize}

\section*{Acknowledgements} I thank the US Department of Energy for
partial support of this research. This talk is based on Ref.~\cite{cc},
and I thank my co-authors, Ron Kantowski, Yun Wang, and Chung Kao,
for their contributions.

\end{document}